# Integrated Architecture for the Automated Generation and Coil Stabilization of a PZT-Enabled Microcomb

Mark Harrington[1*], Steven M. Zhu[1], Rahul Chawlani[1] Kaikai Liu[1], Shuman Sun[2], Ruxuan Liu[2], Xu Yi[2,3]
Daniel J. Blumenthal[1†]

[1]Department of Electrical and Computer Engineering, University of California Santa Barbara, Santa Barbara, CA 93106

[2]Department of Electrical and Computer Engineering, University of Virginia, Charlottesville, VA, USA

[3]Department of Physics, University of Virginia, Charlottesville, VA, USA

Corresponding Author: danb@ucsb.edu
*

(Dated: June 21, 2025)

## ABSTRACT

Silicon nitride integrated Dissipative Kerr Soliton (DKS) microcombs have emerged as a future solution to bring metrological optical frequency comb (OFC) capabilities into a photonic integrated platform with mass-scale fabrication benefits. Precision applications demand low comb line phase noise as well as high repetition rate stability, including quantum sensing and computing, precision metrology, and microwave generation. However, current approaches to reducing comb noise and repetition rate stabilization involve complex architectures, multiple lasers, and high-power components, that combined are challenging to integrate to the chip scale. A CMOS compatible system for on-chip DKS comb ignition and stabilization, capable of providing lab-based performance to non-experts and portable applications, remains to be demonstrated. To achieve this goal, new architectures are needed to simplify the comb control, stabilization, actuation, and pump laser requirements, while lowering the overall power and enabling integrated solutions. Here we demonstrate a greatly simplified stabilized DKS comb architecture with a single laser, and a single point electronic control of both the microcomb generation and its stabilization to a 16-meter-long coil-resonator reference. The PZT controlled microcomb and coil-resonator reference are both fabricated in the low loss CMOS compatible silicon nitride integration platform. PZT-enabled control brings flexibility and simplicity to the soliton generation and comb line frequency noise reduction using a single CW fixed frequency pump laser, resulting in significantly reduced electronic and optical infrastructure. The silicon nitride microcomb is integrated with a low power, broadband, lead-zirconate-titanate (PZT) actuator that is driven by a simple electronic control sequence that establishes both soliton generation and dual wavelength lock to the 16-meter coil resonator. We demonstrate coil-resonator thermorefractive noise (TRN) limited operation with comb line stabilization as low as 66 Hz integral linewidth and repetition rate phase noise of the 108 GHz soliton equivalent to -118 dBc/Hz when divided down to 10 GHz. We also demonstrate that this approach is able to suppress the 1 kHz frequency noise by 4 orders of magnitude, to below 10 $Hz^2$/Hz, over the complete 35 nm wide comb spectrum. The low power PZT actuator consumes nW bias power and the coil resonator allows flexible dual locking using arbitrary comb lines. These results show a clear path towards full chip integration of stabilized soliton microcombs with simplicity and versatility absent in other schemes. Such integrated architectures will open up their use in deployable ultra-low phase noise mmWave and RF generation, fieldable precision metrology, quantum sensing, portable and space-based applications and high-capacity coherent fiber communications.

# 1. INTRODUCTION

Dissipative Kerr Soliton (DKS) microcombs[1–4], have emerged as a promising candidate for integrated multiwavelength optical sources for a wide variety of precision applications including ultra-low phase noise microwave generators[5–10] atomic clocks[11], quantum sensing[12,13] and precision spectroscopy[14–16]. Commercial stabilized fiber frequency combs have served as the workhorse for these precision applications by leveraging ultra-low expansion (ULE) vacuum cavities and self-referencing to provide ultra-low phase noise and exquisite linewidth and stability. Silicon nitride ($Si_3N_4$) DKS microcombs have emerged as a leading integration approach due to the low optical loss, compatibility with CMOS semiconductor foundry processes[17], and potential for higher level system-on-chip integration[18]. Bringing lab-scale stabilized combs to the integrated chip-scale, while providing ease of use, reduced number of components, and low power consumption, will enable a wide range of portable and scalable precision applications. Yet to date, stabilization of DKS microcombs has required complex architectures, multiple lasers, high power actuation, and components that present integration challenges.

Stabilized DKS microcombs experience tradeoffs between ease of use and low phase noise operation and stabilization[19,20]. Improved robustness can be achieved by incorporating discrete modulators and lasers, adding to system size, cost, complexity and power consumption[3,21,22], and the comb generation process can be simplified using self-injection locking (SIL)[23,24]. However, these techniques do not address simplification of comb stabilization which requires actuation of two degrees of freedom independently, namely the soliton repetition rate ($f_{rep}$) and the carrier envelope offset frequency ($f_{ceo}$). To date, stabilization solutions require interaction between the thermorefractive noise (TRN) and the coupling of the modulation methods to both degrees of freedom[21], making integration difficult.

Thermal tuning actuation for robust soliton generation has been shown in the $Si_3N_4$ platform[25], but the slow thermal bandwidth limits the ability to provide stabilization at the same time as well as consuming high power. Piezoelectric modulation based on aluminum nitride (AlN)[26,27] has been used to modulate soliton microcomb at high speed, but it has small tuning range and poses limitation on direct soliton generation. Therefore, New approaches are needed to simplify stabilized DKS combs and provide a path to integration with CMOS compatible platforms.

Here, we demonstrate a greatly simplified frequency stabilized DKS microcomb architecture capable of robust low power soliton generation and repetition rate locking to an integrated meter-scale reference cavity using a single point control sequence and single fixed frequency laser. This automatic single point control generation and stabilization produces ultra-low phase noise for all comb lines and the soliton repetition rate, with noise characteristics transferred from the meter-scale cavity. The architecture is enabled by an integrated Lead Zirconate Titanate (PZT) actuated $Si_3N_4$ microring resonator that is stabilized to an integrated $Si_3N_4$ 12 MHZ free spectral range (FSR) 16-meter-long coil-resonator reference. This approach differs from techniques that rely on mixing two beat notes to decouple the comb pump frequency from the repetition rate, thus stabilizing only soliton repetition frequency[5,6,9,10].

The large mode volume of the 16-meter coil-resonator provides a low TRN floor and soliton stabilization with comb line integral linewidths as low as 66 Hz. We additionally show that the measured frequency noise of the soliton comb lines match existing two-point locked frequency comb noise models[28]. Importantly, the soliton repetition rate, nominally 108 GHz in this work, can be directly tuned to sweep soliton lines through multiple coil-resonator free spectral ranges (FSR) and allow direct error signal generation for comb stabilization. Using this feature, we demonstrate suppression of the 1 kHz comb line frequency noise by 4 orders of magnitude, to below 10 $Hz^2/Hz$ over the complete 35 nm wide comb spectrum. We also measure the soliton repetition rate phase noise to be -25 dBc/Hz at 10 Hz, -75 dBc/Hz at 1 kHz, and -98 dBc/Hz at 10 kHz offset for 108 GHz $f_{rep}$, equivalent to -118 dBc/Hz when divided down to 10 GHz. The PZT actuation provides DC-coupled broadband actuation and preserves the low silicon nitride waveguide loss and high resonator quality factor (Q)[29]. This results in an order of magnitude improvement in linear tuning strength (170 MHz/V) and DC-coupled broadband modulation bandwidth

(DC-70 MHz) compared to AlN microcombs[27], enabling a 10 Volt peak-to-peak control voltage and high speed soliton feedback control. We then show that the same soliton electronic PZT actuation signal can be used to decouple control of both soliton degrees of freedom to realize two-point soliton stabilization. The high speed PZT feedback, combined with the 12 MHz coil FSR, allows us to directly lock both the soliton pump and an arbitrary soliton comb line to two resonances of a 16-meter $Si_3N_4$ coil resonator without modulating the pump laser frequency or amplitude[1,19,21] or requiring a second reference laser[30]. In this manner, the PZT actuated lock provides simplified repetition frequency control that is inherently decoupled from the pump laser frequency, allowing the pump laser to remain locked to the coil reference and stabilize both the soliton repetition frequency and the frequency of each comb line. These results demonstrate that simplified, automatic generation and stabilization of DKS microcombs, integrated using the CMOS compatible silicon nitride platform, have the potential to move these important metrological tools from the laboratory out to the field to realize a wide range of portable precision applications.

## 2. RESULTS

### A. Coil stabilized PZT microcomb

The simplified stabilized DKS architecture, depicted in Fig.1a, is centered around a PZT actuated silicon nitride DKS microcomb driven by a single sequential electronic control sequence that first generates a soliton and then stabilizes the soliton repetition rate to an integrated 16-meter $Si_3N_4$ coil resonator, using only a fixed wavelength laser and single point electronic control. The long coil resonator length results in a low thermorefractive noise (TRN) floor, reducing the low- and mid-range offset frequency noise of the stabilized DKS microcomb. By stabilizing the pump laser at $\nu_p$ and an extracted soliton line at $\nu_s$ to another fringe of the coil resonator, frequency fluctuations in both the soliton repetition rate ($f_{rep}$) and the frequency of any soliton line $\nu_m$ are strongly suppressed. The ~ 12 MHz free-spectral range (FSR) of the 16-meter coil resonator provides a fine frequency locking grid with which to stabilize any of the DKS comb lines through a direct PDH lock, removing the need for a separate local oscillator laser and phase-locked loop and providing flexibility and robustness of the dual lock process.

The PZT microcomb actuator is driven with an initial voltage sequence (top trace Fig. 2a) used to generate a bright soliton (Fig. 2b) pumped by a stationary continuous-wave (CW) pump laser. Then the control sequence is programmed to modulate $f_{rep}$ for the dual point locking step (Fig. 2c). The pump laser frequency is unperturbed by the generation and stabilization process and remains directly PDH-locked to the coil resonator at 1550 nm. Once a soliton has been generated and the pump and microcomb locked to the coil, the subsequent PZT voltage triangle wave modulation (upper trace Fig. 2c) results in a change in laser-resonator detuning as shown in the bottom trace of Fig. 2c, which results in a modulation of $f_{rep}$[21]. Importantly, this modulation is decoupled from the comb's second degree of freedom since modulation of the microring resonance frequency does not affect the pump laser frequency. As optical frequencies of all soliton lines $\nu_m$ are constrained by the relationship $\nu_m = \nu_p + mf_{rep}$, modulation of $f_{rep}$ creates a proportional modulation to every soliton line, which allows for any soliton line $\nu_m$ to be used for direct PDH-locking of a second longitudinal mode of the same coil resonator, an important feature of this architecture. The pump and soliton line are locked to resonator modes that have similar optical mode profiles in the waveguide, leading to strong common mode TRN noise reduction for frep. In addition, access to many resonances within a single soliton range enables fine control of soliton line and repetition frequency by changing the target coil resonance, as well as finer tuning using PDH lock offset, as seen in Fig. 2c. Fig. 2d shows coil resonator transmission and resulting PDH error signals as the pump laser (soliton lines) at 1550 nm (1535 and 1560 nm), are swept across a coil resonance. We choose to lock a soliton line near 1560 nm (m=11) to the coil resonator as a compromise between soliton output power, frequency difference from the pump, and coil resonator quality factor, as discussed in detail later.

Since both degrees of freedom are constrained by the dual-locking, the stability and noise properties of the 16-meter coil resonator is transferred to all other comb lines. It can be shown (See Supplement Sec. III) that the frequency noise $S_{\nu_m}(f)$ of come line $m$ follows the form:

$$S_{\nu_m}(f) = S_{\nu_p}(f)\left(1-\frac{m}{r}\right)^2 + S_{\nu_r}(f)\left(\frac{m}{r}\right)^2 + 2C(f)\sqrt{S_{\nu_p}(f)S_{\nu_r}(f)}\left(1-\frac{m}{r}\right)\frac{m}{r},$$

where $S_{\nu_m}$, $S_{\nu_p}$, and $S_{\nu_r}$ are the frequency noise power-spectral densities (PSDs) of the $m_{th}$ soliton line, the pump laser, and the stabilized soliton line, respectively, $m$ and $r$ are the integer soliton line numbers corresponding to the measured line and stabilized line, and $C(f)$ is the degree of correlation between the noise on the two PDH locks. As shown later in the noise measurements, this results in a significant flattened reduction in the frequency noise across all soliton comb lines. The advantage of this architecture over a soliton locked at only a single point is removal of additional $f_{rep}$ noise from the microcomb TRN and pump RIN[31] that scales as the comb line distance from the pump in $m^2$.

## B. PZT actuated Si$_3$N$_4$ microring

A cross-section of the PZT actuated microresonator waveguide is shown in Fig. 3a. The waveguide geometry has a 2.3 um x 0.8 um Si$_3$N$_4$ core cross-section with a surrounding silicon dioxide (SiO$_2$) upper cladding. The piezo-electric actuator is fabricated on top of the oxide cladding and consists of two concentric rings of 0.5 um thick PZT laterally offset from the waveguide center by 3 um and clad with platinum control electrodes. The linear tuning response is measured to be -170 MHz/V and 110 MHz/V for the TM00 and TE00 modes, respectively (Fig. 3b). The negative TM tuning coefficient denotes a redshift with increasing voltage, while the TE mode is blue shifted with the same applied bias. We find DC tuning is linear from 1-20 V except for a small amount of hysteresis at low tuning voltage[29]. We do not report responses at negative voltages in this work since the actuator is used in a single ended mode. No leakage current was detected for any DC voltages used and is estimated to be less than our measurement floor of 0.1 nA. As a result, we estimate the actuator's power consumption is below 2 nW under DC tuning, while

AC power consumption is dependent on the modulation bandwidth and format and electronic driver design. We measure a modulation bandwidth at 1550 nm for the TM and TE modes to be 70 and 30 MHz respectively (Fig. 3c). The actuator frequency response is flat and remains within +/- 3 dB of the DC response within the reported bandwidth. As the loaded TM (TE) quality factors are 2.2e6 (5.5e6), we find the TE modulation bandwidth is primarily limited by the 35 MHz resonance linewidth while the TM bandwidth is limited by a combination of resonator linewidth, electrical actuation, and PZT optomechanical response[29].

## C. Soliton generation and actuation

Soliton generation is achieved by applying a sequence of voltage ramps to the PZT actuator, which rapidly sweeps the microcomb resonance across the fixed-frequency pump laser. The ramp sequence begins with an initial resonance slew rate of 10 MHz/us, which rapidly tunes the resonator through the blue-detuned side of resonance and into the soliton existence range before thermal effects can take hold. This is followed by longer time sequences of progressively slower resonator tuning that allow the soliton to reach thermal equilibrium. We generate a single soliton with a 9.9Vpp ramp signal, of which 9.8 V is accounted for by the steep initial ramp, corresponding to a total frequency shift of 1.7 GHz which is approximately 20 resonator linewidths. The presence of a single soliton state is confirmed by the narrow repetition frequency spectrum and characteristic $sech^2$ envelope which contains 40 lines (~ 4THz) within a 3dB power envelope (see Fig. 2b). Once the soliton has been initiated, it is then modulated through the same PZT actuator voltage, which with the fixed pump frequency, results in modulation of the laser-resonator detuning and the soliton pulse width and repetition frequency. While the exact tuning strength of $f_{rep}$ varies with the initial pumping conditions of the soliton, it ranges from 1-5 MHz per volt and can tune any soliton line more than a full 12 MHz coil FSR.

## D. Stabilized PZT comb and repetition rate phase noise

We characterize the frequency noise of individual soliton lines as well as the phase noise of $f_{rep}$ using a combination of an optical frequency discriminator (OFD) and heterodyne beatnote detection with a self-referenced fiber comb (Supplementary IV). Fig. 4a shows the measured noise of the two PDH locked frequencies, the pump laser and the 1560 nm soliton line. Both locks demonstrate strong frequency noise suppression close to the TRN noise floor out to 100 kHz. The degree of suppression is currently limited by laser modulation bandwidth on the pump and the optical path delay for the soliton lock, both which can be improved in the future. We measure low reverse $1/\pi$ integral linewidths of 66 Hz for the soliton line and 74 Hz for the pump (orange and blue shaded regions in Fig. 4a). From 10 Hz to 200 Hz, the frequency noise is dominated by common mode environmental perturbations. At 10 kHz frequency offset the pump lock is TRN limited to 0.2 Hz$^2$/Hz (-90 dBc/Hz) and the soliton line lock is limited to 2 Hz$^2$/Hz (-80 dBc/Hz) due to amplified spontaneous emission (ASE), resulting in uncorrelated noise and no common-mode rejection at this frequency.

Next, we measure the frequency noise of other soliton lines that are not involved in the PDH locks and compare the frequency noise of each line at 1 kHz and 10 kHz offsets with and without soliton coil-stabilization. Fig. 4b. shows the frequency noise of dual-locked soliton lines (circles) and soliton lines with only at the pump laser locked (triangles) demonstrating over 4 orders of magnitude noise suppression across the comb bandwidth. The dashed lines in Fig. 4b are generated using the analytical model above in good agreement with measurements. For the analytical curves we apply the frequency noise of the locked pump and soliton line (r = 11) and assume $C(f) = 0$. The strong agreement between analytical and measured results at 10 kHz is most likely due to uncorrelated noise between the two locks. At 1 kHz, we measure comb line noise lower than the $C = 0$ model, indicating that a significant fraction of the noise at this frequency is correlated environmental perturbations. At both 1 kHz and 10 kHz offsets, frequency noise of the comb lines is suppressed significantly compared to the single lock case. We further verify the performance of our architecture and soliton locking in Fig. 4c by measuring the phase noise of the repetition frequency using a self-referenced commercial fiber comb (Supplementary sec. IV). We observe the effects of an optical frequency division with a factor of 11 which results in phase noise of -100 dBc/Hz at 10 kHz offset, -50 dBc/Hz at 100 Hz offset and soliton repetition rate phase noise equivalent to -118 dBc/Hz when divided down to 10 GHz

## 3. DISCUSSION

In this work we present an architecture that greatly simplifies the automatic generation and stabilization of a PZT-enabled DKS microring soliton comb to a large mode volume integrated reference cavity. The architecture only requires a fixed frequency pump laser and a single point control electrical programming sequence for both soliton generation and dual-point locking to the reference cavity, making this approach amenable to photonic integration, greatly reducing the component count, and providing a tool to move stabilized combs out of the lab for non-experts and into portable applications. Both the PZT microcomb and 16-meter coil resonator are fabricated in the low loss silicon nitride CMOS foundry compatible platform. Stabilizing the DKS comb to the coil allows the comb lines to inherit the coil frequency noise properties, leading to over 4 orders magnitude noise reduction at 1 kHz and 10 kHz offset frequencies, to below 10 $Hz^2$/Hz across the entire 35 nm comb bandwidth. This stabilization results in as low as 66 Hz integral linewidth and soliton repetition rate phase noise of -118 dBc/Hz at 10 kHz offset when divided down to 10 GHz. The low voltage and power requirements of the PZT are an order of magnitude improvement over previous stress-optic microcomb modulation approaches, and the 12 MHz FSR enables PZT locking of any comb line in this dual lock configuration, while the entire system is pumped by a single laser with no fast tuning or injection-locking requirements.

Further experiments can be performed to precisely measure the common-mode noise behavior of the two PDH locks, the full effect of soliton PDH wavelength choice on linewidth scaling of soliton lines, as well as the impact of the microring deign on achievable system performance. Furthermore, the noise performance, integration level, and automation of this system are straightforward to advance; division factors and locking quality can be increased by employing a tunable dispersive wave[9] to create high-power comb lines far from the pump wavelength and PDH phase modulation can be achieved through a combination of laser current modulation and a high frequency comb dither signal.

Looking forward, integration of this architecture on a single chip using the silicon nitride platform is illustrated in Fig. 5. Optical amplification can be a eliminated due to minimal interconnection loss and the removal of facet and filtering losses. Multi-layer nitride designs[32] can be used to locate the thick core DKS comb on a top nitride layer and the pump laser and filters and reference cavity on a lower thin nitride layer. Techniques such as an anneal-free ultra-low loss platform have the potential to implement both thin and thick nitride components using the same fabrication process and temperatures limited to 250C[18]. The nitride platform has demonstrated components and operation across the visible to SWIR range[33–37]. These results hold promise to unlock the use of frequency stabilized DKS soliton frequency combs across a wide array of quantum, precision and portable timing, navigation, microwave, and communications applications.

## Acknowledgements

This work is based in part by funding from DARPA GRYPHON award HR0011-22-2-0008 and Army Research Laboratories (ARL) under grant number W911NF-22-2-0056. The views and conclusions contained in this document are those of the author(s) and should not be interpreted as representing the official policies of DARPA or the U.S. government. The authors gratefully acknowledge measurement automation provided by Joshua Hansen, and helpful discussion with Gregory Moille and Kartik Srinivasan.

## Author contributions

M.W.H., D.J.B, and X.Y. conceived the work. M.W.H. operated the system and analyzed results with the assistance of S.M.Z. and R.C.. S.M.Z. packaged, characterized, and analyzed the photonics with assistance from M.W.H, R.C., and K.L.. All authors discussed the results and contributed to the writing of the paper. D.J.B. and X.Y. supervised the research.

## Competing Interests

The Authors declare no competing interests.

## Data availability

The data that support the plots within this paper and other finding of this study are available from the corresponding author upon reasonable request.

## Code availability

The codes that support the findings of this study are available from the corresponding authors upon request.

## Correspondence

Correspondence and requests for materials should be addressed to D.J.B. (danb@ucsb.edu).

## METHODS

### Soliton line frequency noise and integral linewidth

We directly measure the power spectral density (PSD) of frequency fluctuations in the soliton pump laser and generated comb lines using a combination of optical frequency discriminator and commercial fiber comb referencing. The optical frequency discriminator method uses an unbalanced Mach Zender interferometer (MZI) with 200m delay length to convert laser frequency fluctuations into photodetected amplitude fluctuations, as described in more detail in other works[38]. This method is accurate at frequency offsets greater than 2 kHz, as noise in the fiber MZI dominates the measurement at lower frequencies. For lower frequency offsets, absolute noise on the pump laser and locked soliton line were measured via a heterodyne beat-note with a cavity-stabilized fiber frequency comb. The frequency comb is self-referenced and optically referenced to a Hz-level ultra-low-expansion (ULE) cavity-locked laser. Fluctuations in the beat-note are recorded with a frequency counter and phase noise analyzer and stitched with the OFD measurement to obtain a complete frequency noise measurement. Integral linewidth of these lasers is calculated from frequency noise PSD using the integration of phase noise method[38], where laser phase noise is integrated from frequency offset of infinity to the frequency at which the integral equals $1/\pi$ rad$^2$.

## Soliton line f$_{rep}$ noise measurement

The PSD of frequency fluctuations in the soliton repetition frequency (f$_{rep}$) are measured indirectly by analyzing relative frequency fluctuations between the soliton pump and locked soliton line optical frequencies. This is done by changing the optical reference of the fiber comb from the ULE cavity to the coil-stabilized soliton pump laser. We then measure the fluctuations of the heterodyne beat-note between the pump-referenced fiber comb and the locked soliton line at 1560 nm. As the fiber comb is still self-referenced, comb lines in the vicinity of the locked soliton line take on the frequency fluctuations of the soliton pump, and the heterodyne therefore compares the difference of fluctuations of the soliton line and pump laser. Frequency and phase noise of f$_{rep}$ are then obtained by dividing the PSD of the relative fluctuations by the square of the soliton line mode number (11 in this case).

## PZT actuation measurement

PZT tuning strength of the microresonator is measured by progressively changing the voltage applied to the PZT while recording the resulting resonance frequency change. Resonance frequency shift is measured using a laser sweep calibrated by an unbalanced Mach Zender interferometer with 185 MHz FSR. The measurement sequence is automated to reduce total measurement time below 5 seconds per point to minimize errors due to frequency drift in the laser, microring and MZI. Electro-optic modulation bandwidth is measured using a network analyzer – electrical stimulus is routed to the PZT actuator while a laser is manually manually held at the half-maximum of the 1550 nm resonance dip. The optical power passing through the resonator is sent to a 25 GHz amplified photodetector, and the resulting RF signal is measured by the network analyzer.

**PZT Soliton Generation**

We generate stable solitons with a stationary pump laser through application of a voltage signal composed of various slopes, as described above. This signal is generated by an arbitrary waveform generator (AWG) and amplified with a high-speed operational amplifier to enable drive voltages of up to +15VDC to the positive PZT electrode. Once the soliton is generated, the AWG and op-amp are isolated from the PZT by a mechanical switch, which smoothly disconnects the AWG source and connects a low-noise DC power supply set to the final voltage of the generation signal. Solitons were generated in solitons operating from both TE00 and TM00 polarization modes, with similar drive voltage requirements but opposite ramp slopes ramps. Soliton modulation is driven from the negative PZT electrode to avoid use of voltage summing electronics.

**REFERENCES**

1. Herr, T. *et al.* Temporal solitons in optical microresonators. *Nature Photon* **8**, 145–152 (2014).

2. Yi, X., Yang, Q.-F., Yang, K. Y., Suh, M.-G. & Vahala, K. Soliton frequency comb at microwave rates in a high-Q silica microresonator. *Optica, OPTICA* **2**, 1078–1085 (2015).

3. Brasch, V. *et al.* Photonic chip–based optical frequency comb using soliton Cherenkov radiation. *Science* **351**, 357–360 (2016).

4. Gaeta, A. L., Lipson, M. & Kippenberg, T. J. Photonic-chip-based frequency combs. *Nature Photon* **13**, 158–169 (2019).

5. Sun, S. *et al.* Integrated optical frequency division for microwave and mmWave generation. *Nature* **627**, 540–545 (2024).

6. Kudelin, I. *et al.* Photonic chip-based low-noise microwave oscillator. *Nature* **627**, 534–539 (2024).

7. Zhao, Y. *et al.* All-optical frequency division on-chip using a single laser. *Nature* **627**, 546–552 (2024).

8. Sun, S. *et al.* Microcavity Kerr optical frequency division with integrated SiN photonics. *Nat. Photon.* 1–6 (2025) doi:10.1038/s41566-025-01668-3.

9. Ji, Q.-X. *et al.* Dispersive-wave-agile optical frequency division. *Nat. Photon.* 1–6 (2025) doi:10.1038/s41566-025-01667-4.

10. Jin, X. *et al.* Microresonator-referenced soliton microcombs with zeptosecond-level timing noise. *Nat. Photon.* 1–7 (2025) doi:10.1038/s41566-025-01669-2.

11. Newman, Z. L. *et al.* Architecture for the photonic integration of an optical atomic clock. *Optica, OPTICA* **6**, 680–685 (2019).

12. Dalvit, D. A. R. *et al.* Quantum Frequency Combs with Path Identity for Quantum Remote Sensing. *Phys. Rev. X* **14**, 041058 (2024).

13. Sun, Y. *et al.* Applications of optical microcombs. *Adv. Opt. Photon., AOP* **15**, 86–175 (2023).

14. Kalashnikov, V. L. & Sorokin, E. Soliton absorption spectroscopy. *Phys. Rev. A* **81**, 033840 (2010).

15. Suh, M.-G., Yang, Q.-F., Yang, K. Y., Yi, X. & Vahala, K. J. Microresonator soliton dual-comb spectroscopy. *Science* **354**, 600–603 (2016).

16. Yang, Q.-F. *et al.* Vernier spectrometer using counterpropagating soliton microcombs. *Science* **363**, 965–968 (2019).

17. Moss, D. J., Morandotti, R., Gaeta, A. L. & Lipson, M. New CMOS-compatible platforms based on silicon nitride and Hydex for nonlinear optics. *Nature Photon* **7**, 597–607 (2013).

18. Bose, D. *et al.* Anneal-free ultra-low loss silicon nitride integrated photonics. *Light Sci Appl* **13**, 156 (2024).

19. Brasch, V., Geiselmann, M., Pfeiffer, M. H. P. & Kippenberg, T. J. Bringing short-lived dissipative Kerr soliton states in microresonators into a steady state. *Opt. Express, OE* **24**, 29312–29320 (2016).

20. Carmon, T., Yang, L. & Vahala, K. J. Dynamical thermal behavior and thermal self-stability of microcavities. *Opt. Express, OE* **12**, 4742–4750 (2004).

21. Stone, J. R. *et al.* Thermal and Nonlinear Dissipative-Soliton Dynamics in Kerr-Microresonator Frequency Combs. *Phys. Rev. Lett.* **121**, 063902 (2018).

22. Xiao, Y. *et al.* Optimizing auxiliary laser heating for Kerr soliton microcomb generation. *Opt. Lett., OL* **49**, 1129–1132 (2024).

23. Shen, B. *et al.* Integrated turnkey soliton microcombs. *Nature* **582**, 365–369 (2020).

24. Stern, B., Ji, X., Okawachi, Y., Gaeta, A. L. & Lipson, M. Battery-operated integrated frequency comb generator. *Nature* **562**, 401–405 (2018).

25. Joshi, C. *et al.* Thermally controlled comb generation and soliton modelocking in microresonators. *Opt. Lett., OL* **41**, 2565–2568 (2016).

26. Tian, H. *et al.* Hybrid integrated photonics using bulk acoustic resonators. *Nat Commun* **11**, 3073 (2020).

27. Liu, J. *et al.* Monolithic piezoelectric control of soliton microcombs. *Nature* **583**, 385–390 (2020).

28. Moille, G., Shandilya, P., Stone, J., Menyuk, C. & Srinivasan, K. All-Optical Noise Quenching of An Integrated Frequency Comb. Preprint at https://doi.org/10.48550/arXiv.2405.01238 (2025).

29. Wang, J., Liu, K., Harrington, M. W., Rudy, R. Q. & Blumenthal, D. J. Silicon nitride stress-optic microresonator modulator for optical control applications. *Opt. Express* **30**, 31816 (2022).

30. Zhou, H. *et al.* Soliton bursts and deterministic dissipative Kerr soliton generation in auxiliary-assisted microcavities. *Light Sci Appl* **8**, 50 (2019).

31. Lei, F. *et al.* Optical linewidth of soliton microcombs. *Nat Commun* **13**, 3161 (2022).

32. John, D. D. *et al.* Multilayer Platform for Ultra-Low-Loss Waveguide Applications. *IEEE Photonics Technology Letters* **24**, 876–878 (2012).

33. Liu, K. *et al.* Tunable broadband two-point-coupled ultra-high-*Q* visible and near-infrared photonic integrated resonators. *Photon. Res., PRJ* **12**, 1890–1898 (2024).

34. Heim, D. A. S., Bose, D., Liu, K., Isichenko, A. & Blumenthal, D. J. Photonic Integrated External Cavity Coil-Resonator Stabilized Laser with Hertz-Level-Fundamental and Sub-250-Hertz Integral Linewidth. in *2024 Conference on Lasers and Electro-Optics (CLEO)* 1–2 (2024).

35. Chauhan, N. *et al.* Ultra-low loss visible light waveguides for integrated atomic, molecular, and quantum photonics. *Opt. Express, OE* **30**, 6960–6969 (2022).

36. Nejadriahi, H. *et al.* Sub-100 Hz intrinsic linewidth 852 nm silicon nitride external cavity laser. *Optics Letters, Vol. 49, Issue 24, pp. 7254-7257* (2024) doi:10.1364/OL.543307.

37. Huffman, T., Baney, D. & Blumenthal, D. J. High Extinction Ratio Widely Tunable Low-Loss Integrated Si3N4 Third-Order Filter. *arXiv preprint arXiv:1708.06344* (2017).

38. Brodnik, G. M. *et al.* Optically synchronized fibre links using spectrally pure chip-scale lasers. *Nat. Photon.* **15**, 588–593 (2021).

**Figures:**

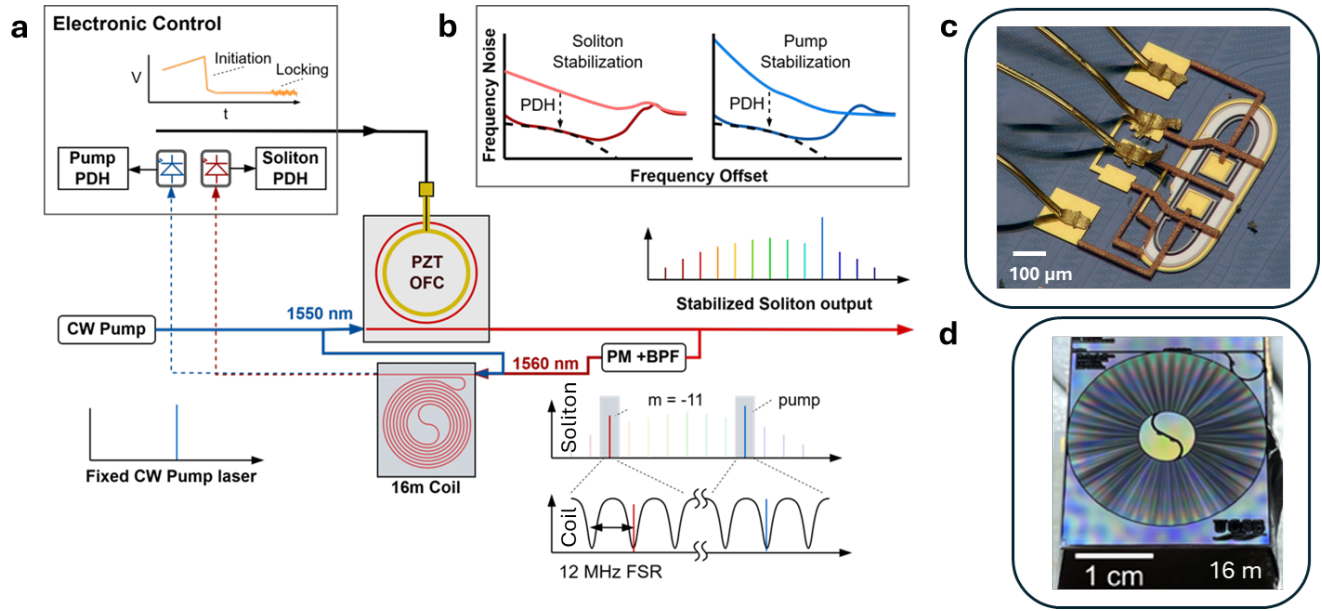

FIG. 1. **Simplified soliton initiation and coil-resonator stabilization via single-point PZT actuated microcomb**. Optical layout of single pump laser and single point electronic control for two-point coil stabilized soliton. The pump laser and a soliton comb line are simultaneously locked via Pound-Drever-Hall (PDH) lock to a low noise $Si_3N_4$ coil resonator. The soliton is initiated through a fast sweep of the PZT without laser tuning, and once initiated, the PZT actuator is then used to lock the soliton line through modulation of the soliton repetition frequency. **b.** Optical micrograph of the PZT-actuated SiN microcomb. **c**. Image of the 16m $Si_3N_4$ coil resonator used as the two-point locking reference. The 12 MHz coil resonator free spectral range enables direct simultaneous locking of the pump laser and any soliton line to the resonator**.** CW; continuous-wave, PM; phase modulator, BPF; Bandpass filter, FSR; Free spectral range

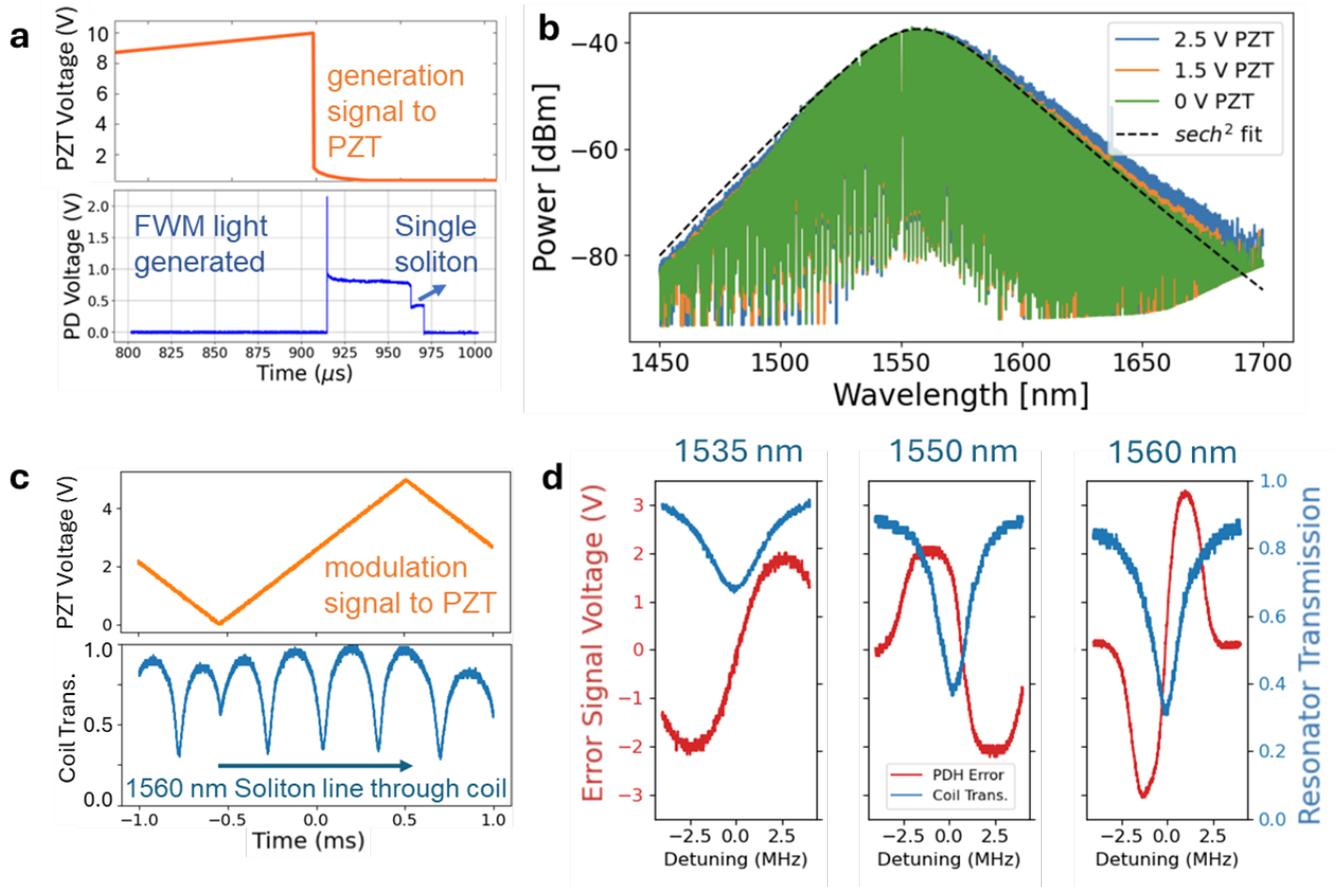

FIG. 2. **PZT actuated Soliton Comb generation and Coil Stabilization. a,** PZT voltage signal (top) and generated four-wave mixing light (bottom) during a soliton initiation cycle. **b,** Optical spectrum of the resulting soliton with a sech$^2$ profile, showing the effect of DC voltage changes applied to the PZT after soliton generation. **c,** PZT voltage ramp signal (top) and of a 1560 nm soliton line through the coil resonator, demonstrating the capability to sweep soliton lines through multiple coil free-spectral ranges. **d**, Coil resonator transmission and Pound-Drever-Hall error signal as soliton lines (1535 and 1560 nm) and pump laser (1550 nm) are frequency swept across a resonance.

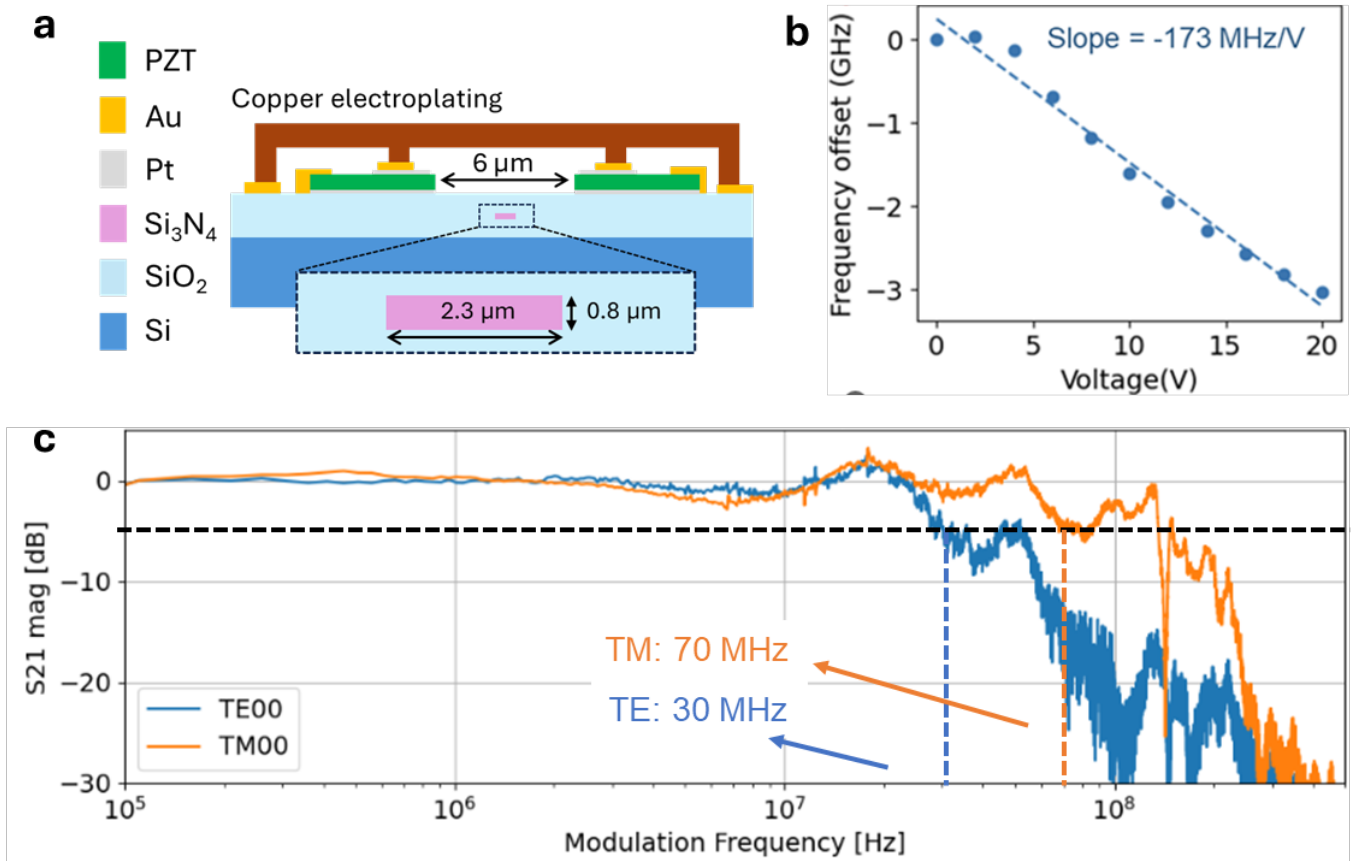

FIG. 3. **PZT soliton generation and modulation. a** Cross-Section of the PZT-actuated microcomb. PZT actuators are offset 3 μm horizontally from the center of the waveguide. **b,** DC frequency shift measurement of TM00 and TE00 modes, respectively. Frequency shift is measured using laser sweeps calibrated by an unbalanced Mach Zender interferometer with 185 MHz FSR. Measurement voltages are sweep automatically at a rate much faster than the observed drift of the measurement laser. **c,** PZT modulation bandwidth for TE and TM modes.

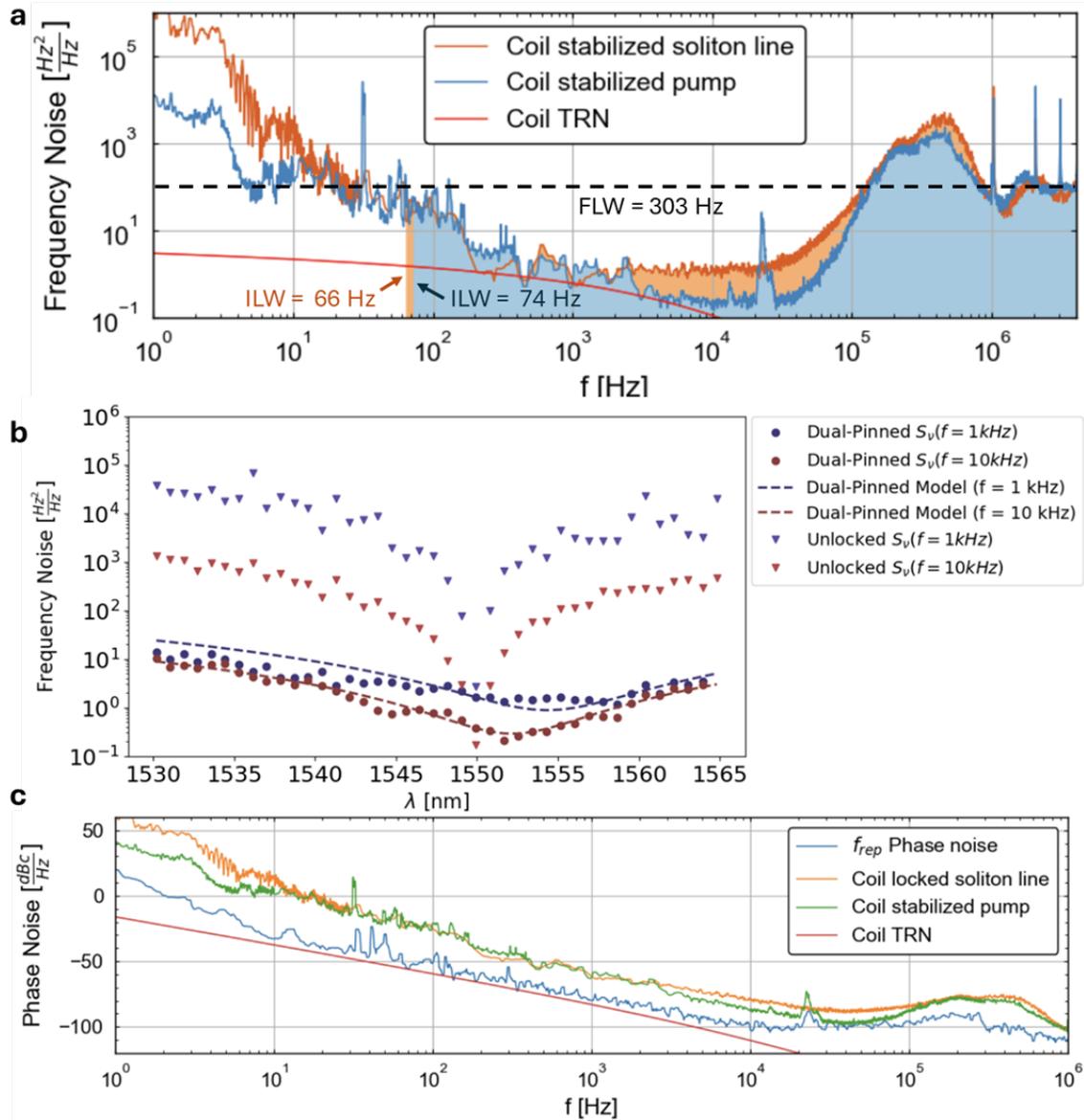

FIG. 4. **Soliton line Frequency noise propagation: a,** Frequency noise power-spectral density of the pump and soliton line locked to the coil resonator shading underneath each trace shows range of frequency noise that integrates to $1/\pi$ rad^2. **b,** Frequency noise of each C-band comb line at 1 kHz and 10 kHz frequency offset. Triangles represent measurements taken while only the pump laser is locked to the coil, while circles represent measurements taken while with pump and the 1560 nm soliton are Pound-Drever-Hall locked. Dashed lines represent the noise level modeled on noise measurements of the two locked points, under the assumption that locked noise is completely uncorrelated. **c,** Phase noise spectrum of the dual-locked soliton repetition frequency ($f_{rep}$) compared to the phase noise of the individual locked points. The red line represents the coil thermorefractive noise limit (TRN) scaled by the optical division factor of $11^2$.

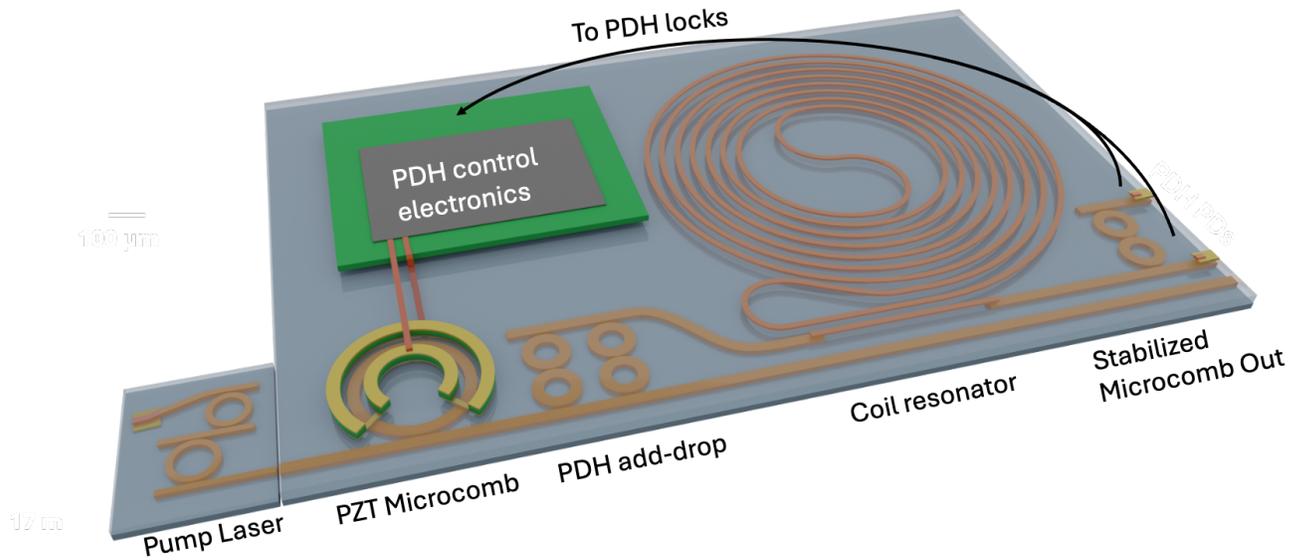

FIG. 5. **Path towards simplified microcomb stabilization**. Vision of a monolithically integrated stabilized microcomb with absolute two-point locking to a stable frequency reference. All necessary photonics components, including microcomb, filters, and coil resonators have been demonstrated in $Si_3N_4$, and can be integrated in a multilayer chip. Soliton initiation and modulation allow for a flexible choice of pump laser and remove discrete components typically required for locking both frequency comb degrees of freedom.

# INTEGRATED ARCHITECTURE FOR THE AUTOMATED GENERATION AND COIL STABILIZATION OF A PZT-ENABLED MICROCOMB: SUPPLEMENTAL DOCUMENT

## I. Large Mode Volume, Low FSR Coil Resonator

Our pump laser and soliton are stabilized to a 16-meter-long coil resonator cavity in deposited silicon nitride on oxide on a 21 mm by 21 mm footprint and 80 nm x 6 $\mu m$ waveguide geometry. This resonator has a low FSR of 12 MHz, an intrinsic quality factor (Q) of 95 million and loaded Q of 76 million at 1550 nm, and an intrinsic Q of 110 million and loaded Q of 89 million at 1560 nm. The chip is fiber packaged in a metal housing for acoustic isolation, with an external heater placed in the housing for long-term temperature stability.

## II. Dual-Wavelength PDH Locking to a Coil Resonator

In this section, we analyze the stability limits of our dual-wavelength coil locking scheme, and compare frequency noise performance to that of the dual-locked soliton. To find the noise penalty associated with the direct PDH-locking of a soliton line to the coil, we construct a similar setup using two independent lasers with no soliton. As in the soliton dual-lock system, we use a counter-propagating setup to increase signal isolation between both wavelengths. Below we show a schematic of the setup.

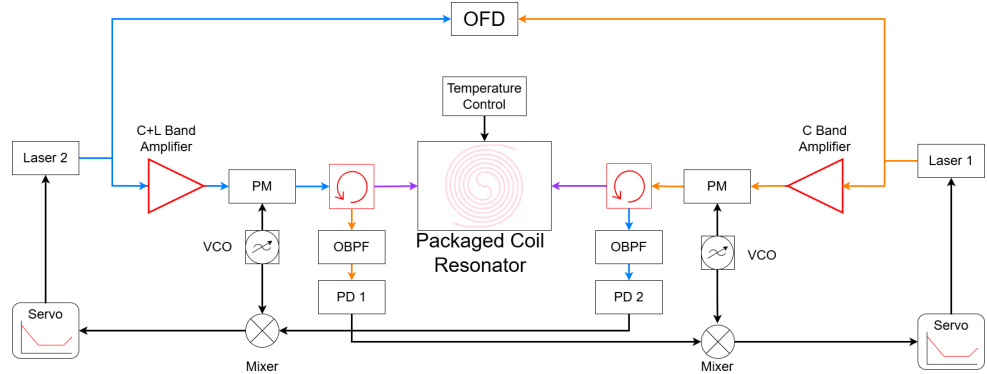

*Fig. S1: Operating Schematic of our Dually Locked Coil Resonator Lock. Counter-propagating optical signals generated by two lasers at either the same or different wavelengths are PDH locked to the packaged coil resonator, which is temperature controlled for long term stability. Both lasers are split and one laser at a time can have its frequency noise characteristic measured by OFD setup. Laser 1 continuous tuning range from 1520-1570 nm. Laser 2 continuous tuning ranges from 1550-1630 nm. Orange: Optical signal corresponding to laser 1. Blue: Optical Signal corresponding to laser 2. Purple: Combined optical signals. Black: Electrical Connections. PM: Phase Modulator, PD: Photodetector, VCO: Voltage Controlled Oscillator, OBPF: Optical Band-Pass Filter, OFD: Optical Frequency Discrimination.*

Here, we discuss the setup with two lasers to better understand the wavelength range of the system. We use two Continuously Tunable Laser (CTLs) where one laser can sweep from 1520-1570 nm (CTL 1) and the other laser (CTL 2) can sweep from 1550-1630 nm (CTL 1: TLB-6728-P, CTL 2: TLB-6730-P-D). CTL 1 is amplified with a C-band EDFA (Erbium Doped Fiber Amplifier) while CTL 2 is amplified by a C+L band EDFA. Modulation sidebands are generated with phase modulators and put through a circulator to demultiplex forward and backward propagating signals. We use the same packaged coil with temperature control described in the previous section. A VCO controls the dither signal and is mixed with our detected signal (Thorlabs PDB-470C) and an error signal and ramp are generated with a servo control module (Vescent D2-125-PL). We use an optical frequency discriminator (OFD) setup using a fiber Mach-Zehnder Interferometer (MZI) to determine the frequency noise at various frequency offsets seen below. Furthermore, we place this setup inside an enclosure with acoustic isolation.

In the figure below, we look at the frequency noise performance at a variety of wavelengths spanning from 1560 nm to 1620 nm when two lasers are counter-propagating in the cavity.

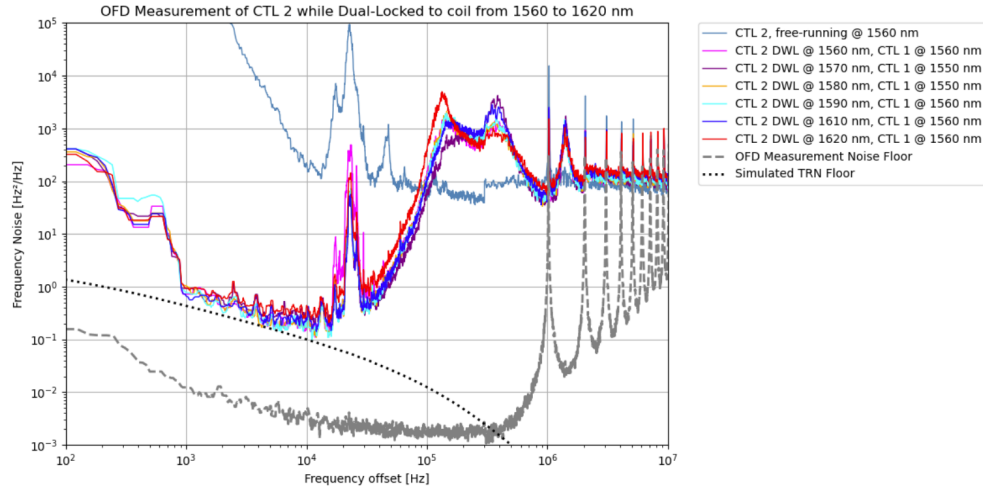

Fig. S2: The figure above shows frequency noise performance across a variety of wavelengths when both lasers are locked to the cavity. They are plotted against the measurement noise floor and thermo-refractive noise floor. OFD: Optical Frequency Discrimination. TRN: Thermo-Refractive Noise. CTL: External Cavity Tunable Laser. DWL: Dual-Wavelength Lock.

As we see, we reach near TRN floor limited performance from 1560-1610 nm with no significant deterioration between the two locks. Comparing the 1560 nm laser lock with our soliton line lock at 1560 nm shows that the performance degradation in the soliton locked noise cannot be explained by the dual-PDH lock system alone. As seen in Supplementary Fig. 3a below, the 1550 nm soliton pump lock (blue) behaves close to the laser-locking case, and resulting unlocked soliton noise at 1560 nm is similar to or less than the noise of an unlocked laser. However, when the soliton line lock is engaged in Supplementary Fig. 3b, we observe an elevated white frequency noise floor on the locked soliton line (red) trace. We observe the locked frequency noise minimum is increased by an amount dependent on the optical signal-to-noise ratio (OSNR) of the source laser and the linewidth of the resonator. This suggests that the soliton line lock frequency noise is limited by ASE contamination of the error signal, which is then scaled by the resonator's discriminator coefficient. This limitation informs the optimal choice of soliton locking line based on a combination of coil resonator linewidth and OSNR. While increasing the mode separation between pump and soliton locks improves $f_{rep}$ noise through a larger division factor, we find that the reduced soliton line power and increased resonator linewidth at this wavelength result in significant noise penalties that outweigh any other benefit. As a result, we find the optimal soliton line choice is located at the comb output maximum, around 1560 nm.

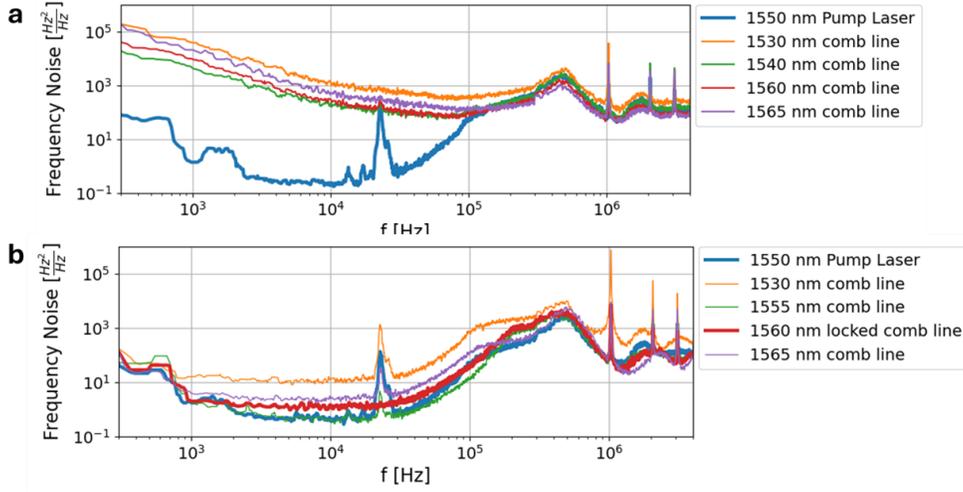

Fig. S3: **a**, Frequency noise spectra of various soliton lines with stabilized pump only, showing rapid degradation in noise levels compared to the pump laser. **b**, Frequency noise spectra of various soliton lines with two-point stabilization, demonstrating that absolute frequency noise of soliton lines showing only small noise increases proportional to µ^2.

## III.    Dual-locked Soliton Noise Propagation

Phase locked solitons always satisfy the following due to conservation of energy:

$$\nu_m = \nu_p + m f_{rep} \tag{1}$$

where m is the soliton mode number relative to the pump mode. As a result, mode-locked frequency combs have only 2 degrees of freedom which are subject to fluctuation, $\nu_p$ and $f_{rep}$. Adding frequency fluctuations on both degrees of freedom yields

$$\bar{\nu}_m + \delta\nu_m = \nu_m, \qquad \bar{f}_{rep} + \delta f_{rep} = f_{rep} \tag{2}$$

Which results in:

$$\bar{\nu}_m + \delta\nu_m = \bar{\nu}_p + \delta\nu_p + m\left(\bar{f}_{rep} + \delta f_{rep}\right) \tag{3}$$

From (1), this simplifies to:

$$\delta\nu_m = \delta\nu_p + m\delta f_{rep} \tag{4}$$

The same relationship applies to the comb line at mode index r, which is stabilized to an external reference $\nu_r$:

$$\delta\nu_r = \delta\nu_p + r\delta f_{rep} \tag{5}$$

Rearranging, we find:

$$\frac{\delta\nu_r - \delta\nu_p}{r} = \delta f_{rep} \tag{6}$$

Therefore, if we control frequency fluctuations on both $\nu_p$ and $\nu_r$ through PDH locking to a coil resonator, we control both degrees of fredom and all other comb line frequencies $\nu_m$ are fully constrained. The relation between fluctuations on the stabilized pump comb line to the fluctiuations on an arbitrary comb line $\nu_m$ can be found via substitution:

$$\delta\nu_m = \delta\nu_p + m\frac{\delta\nu_r - \delta\nu_p}{r} = \delta\nu_p\left(1 - \frac{m}{r}\right) + \delta\nu_r\frac{m}{r} \tag{7}$$

If fluctuations are expressed in the frequency domain, Frequency noise is related to frequency fluctuations by $S_\nu = \|\delta \nu\|^2$. Therefore,

$$S_{\nu_m} = \|\delta \nu_m\|^2 = \left\| \delta \nu_p \left(1 - \frac{m}{r}\right) + \delta \nu_r \frac{m}{r} \right\|^2 \tag{8}$$

If $\delta \nu_p$ and $\delta \nu_r$ are uncorrelated ($C = 0$), then their respective noise terms add in quadrature, and we find:

$$S_{\nu_m} = S_{\nu_p} \left(1 - \frac{m}{r}\right)^2 + S_{\nu_r} \left(\frac{m}{r}\right)^2 \tag{9}$$

However, if $\delta \nu_p$ and $\delta \nu_r$ are possitively correlated ($C = 1$), then their respective noise terms add in phase:

$$S_{\nu_m} = \left( \sqrt{S_{\nu_p}} \left(1 - \frac{m}{r}\right) + \sqrt{S_{\nu_r}} \frac{m}{r} \right)^2 = S_{\nu_p} \left(1 - \frac{m}{r}\right)^2 + S_{\nu_r} \left(\frac{m}{r}\right)^2 + 2\sqrt{S_{\nu_p} S_{\nu_r}} \left(1 - \frac{m}{r}\right) \frac{m}{r} \tag{10}$$

And if $\delta \nu_p$ and $\delta \nu_r$ are negatively correlated ($C = -1$), then their respective noise terms add out of phase:

$$S_{\nu_m} = \left( \sqrt{S_{\nu_p}} \left(1 - \frac{m}{r}\right) - \sqrt{S_{\nu_r}} \frac{m}{r} \right)^2 = S_{\nu_p} \left(1 - \frac{m}{r}\right)^2 + S_{\nu_r} \left(\frac{m}{r}\right)^2 - 2\sqrt{S_{\nu_p} S_{\nu_r}} \left(1 - \frac{m}{r}\right) \frac{m}{r} \tag{11}$$

We can generalize this equation to account for an arbitrary correlation factor:

$$S_{\nu_m} = \left( \sqrt{S_{\nu_p}} \left(1 - \frac{m}{r}\right) + \sqrt{S_{\nu_r}} \frac{m}{r} \right)^2 = S_{\nu_p} \left(1 - \frac{m}{r}\right)^2 + S_{\nu_r} \left(\frac{m}{r}\right)^2 + 2C\sqrt{S_{\nu_p} S_{\nu_r}} \left(1 - \frac{m}{r}\right) \frac{m}{r} \tag{12}$$

We can determine which the degree of correlation $C$ as a function of offset frequency $f$ by comparing absolute frequency noise measurements of $\nu_p$ and $\nu_r$ ($S_{\nu_p}(f)$ and $S_{\nu_r}(f)$), with a measurement of their relative frequency noise $S_{\nu_p - \nu_r}(f)$:

$$S_{\nu_p - \nu_r}(f) = \|\delta \nu_p(f) - \delta \nu_r(f)\|^2 = S_{\nu_p}(f) + S_{\nu_r}(f) - 2C(f)\sqrt{S_{\nu_p}(f) S_{\nu_r}(f)} \tag{13}$$

In the context of this experiment, correlated noise occurs when the locked noise is dominated by coil resonator noise, as any fluctuations the coil length propagate with similar scale factors to resonances at both wavelengths.

### IV. Repetition Rate Stability Measurement Setup

Given that the FSR of our cavity resonance is 108 GHz, most conventional electronics cannot directly measure near-THz level electronic signals. Thus, we use a self-referenced, octave-spanning fiber frequency comb (Vescent FFC-100) with a repetition rate of 100 MHz to measure the soliton's beat note. The fiber comb has internal stabilization of both the repetition rate and the carrier-envelope offset. Specifically, the soliton pump will act as the fiber comb's optical reference. An experimental setup is seen below.

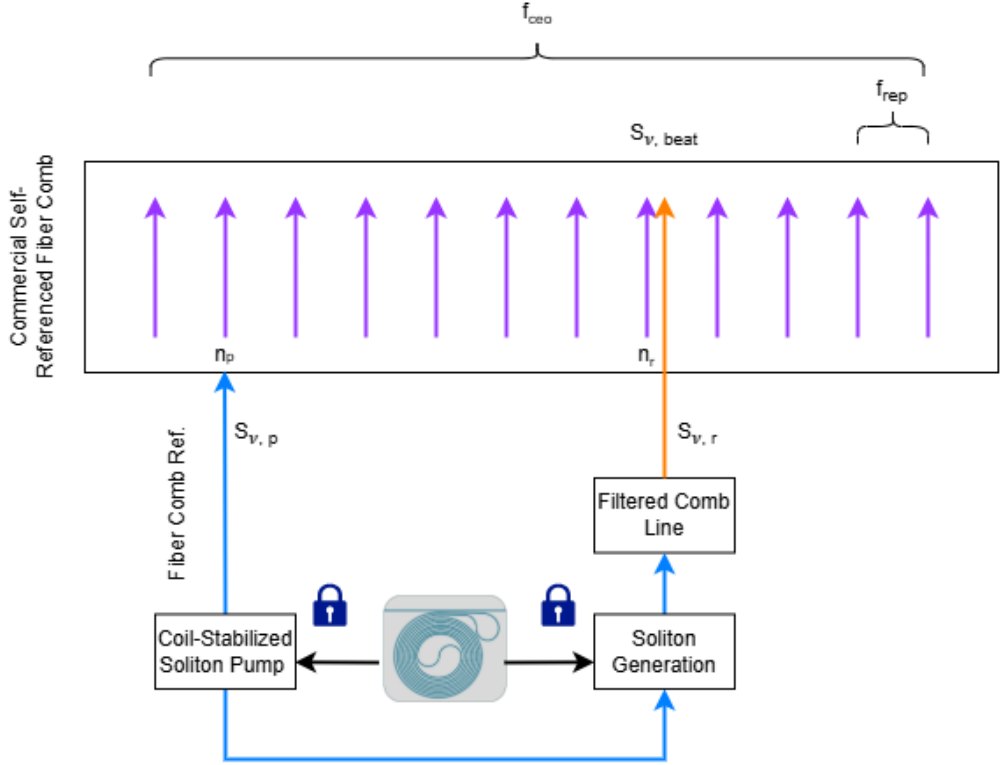

Fig. S4: The figure above provides an overview of our repetition rate stability measurement scheme. In this scheme, a coil-stabilized laser pumps the DKS and acts as the reference for the commercial fiber frequency comb. The comb is fully self-referenced via internal CEO and repetition rate locks. A filtered soliton line from the DKS is heterodyned against a self-referenced comb line for measurement.

A beatnote of the soliton comb line and the nearest fiber comb line gives the noise of the soliton's repetition rate given by the following equation:

$$S_{\nu_{\text{beat}}} = S_{\nu_p}(f) + \left(\frac{n_r}{n_p}\right)^2 S_{\nu_r}(f) \qquad (14)$$

Specifically, by knowing the frequency noise of the beat note and the division factor, the repetition rate noise can be found due to the relative noise between the soliton comb line and the fiber comb line. Specifically, with our pump at 1550 nm and soliton line at 1560 nm, our division factor is .9872.